# HIGH SPEED AND AREA EFFICIENT 2D DWT PROCESSOR BASED IMAGE COMPRESSION

Sugreev Kaur[1] and Rajesh Mehra[2]

[1]ME Student of ECE Department, National Institute of Technical Teachers' Training & Research, Chandigarh, India
`Sugreev.kaur@gmail.com`
[2]Faculty of ECE Department, National Institute of Technical Teachers' Training & Research, Chandigarh, India
`rajeshmehra@yahoo.com`

## ABSTRACT

*This paper presents a high speed and area efficient DWT processor based design for Image Compression applications. In this proposed design, pipelined partially serial architecture has been used to enhance the speed along with optimal utilization and resources available on target FPGA. The proposed model has been designed and simulated using Simulink and System Generator blocks, synthesized with Xilinx Synthesis tool (XST) and implemented on Spartan 2 and 3 based XC2S100-5tq144 and XC3S500E-4fg320 target device. The results show that proposed design can operate at maximum frequency 231 MHz in case of Spartan 3 by consuming power of 117mW at 28 degree/c junction temperature. The result comparison has shown an improvement of 15% in speed.*

## KEYWORDS

DCT, DFT, DWT, JPEG, FPGA.

## 1. INTRODUCTION

With the increasing use of multimedia technologies, image compression requires higher performance. To address needs and requirements of multimedia and internet applications, many efficient image compression techniques, with considerably different features, have been developed [1]. Traditionally, image compression adopts discrete cosine transform (DCT) in most situations which possess the characteristics of simpleness and practicality. DCT has been applied successfully in the standard of JPEG, MPEGZ, etc. However, the compression method that adopts DCT has several shortcomings that become increasing apparent. One of these shortcomings is obvious blocking artifact and bad subjective quality when the images are restored by this method at the high compression ratios [2]. In recent years, many studies have been made on wavelets. An excellent overview of what wavelets have brought to the fields as diverse as biomedical applications, wireless communications, computer graphics or turbulence. Image compression is one of the most visible applications of wavelets. The rapid increase in the range and use of electronic imaging justifies attention for systematic design of an image compression system and for providing the image quality needed in different applications [3].

In recent times, much of the research activities in image coding have been focused on the DWT, which has become a standard tool in image compression applications because of their data reduction capability. In a wavelet compression system, the entire image is transformed and compressed as a single data object rather than block by block as in a DCT-based compression system [4]. It allows a uniform distribution of compression error across the entire image. DWT offers adaptive spatial-frequency resolution (better spatial resolution at high frequencies and better frequency resolution at low frequencies) that is well suited to the properties of an HVS. It can provide better image quality than DCT, especially on a higher compression ratio [5]. Traditionally, Fourier transforms have been utilized for signal analysis & reconstruction. However, Fourier transform does not include any local information about the original signal. Therefore, Short Time Fourier Transform (STFT or Gabor transform) has been introduced, which uniformly samples the time-frequency plane. Unlike the STFT which has a constant resolution at d times and frequencies, the wavelet transform has *a* good time and poor frequency resolution *at* high frequencies, and good frequency and poor time resolution at low frequencies [6]. In JPEG 2000, Discrete Wavelet Transform is used as a core technology to compress still images. It is multi-resolution analysis and it decomposes images into wavelet coefficients and scaling function. In Discrete Wavelet Transform, signal energy concentrates to specific wavelet coefficients. This characteristic is useful for compressing images [7].The multiresolution nature of the discrete wavelet transform is proven as a powerful tool to represent images decomposed along the vertical and horizontal directions using the pyramidal multiresolution scheme. Discrete wavelet transform helps to test different allocations using subband coding, assuming that details at high resolution and diagonal directions are less visible to the human eye. By using an error correction method that approximates the reconstructed coefficients quantization error, we minimize distortion for a given compression rate at low computational cost. The main property of DWT is that it includes neighbourhood information in the final result, thus avoiding the block effect of DCT transform. It also has good localization and symmetric properties, which allow for simple edge treatment, high-speed computation, and high quality compressed image [8]. The 2D DWT has also gained popularity in the field of image and video coding, since it allows good complexity-performance tradeoffs and outperforms the discrete cosine transform at very low bit rates [9]. In general the wavelet transform requires much less hardware to implement than Fourier methods, such as the DCT.

Because of the popularity of wavelets, it is imperative to explore hardware implementations. Hardware concepts such as pipelining and distributed arithmetic may help achieve better throughput [10]. Recent advances in FPGA technology not only provides significant increase in the logic real estates, but also furnishes relatively significant amount of flexible internal RAM modules. An efficient implementation of Discrete Wavelet Transform (DWT) in JPEG2000 is designed with low memory and high pipeline architecture. The corresponding line-based FPGA lifting scheme is put forward from hardware perspective [11]. So in this paper, an area efficient design of 2D DWT Processor for image compression is designed and implemented on an FPGA device.

## 2. DISCRETE WAVELET TRANSFORM

The Discrete Wavelet Transform, which is based on sub-band coding, is found to yield a fast computation of Wavelet Transform. It is easy to implement and reduces the computation time and resources required. The discrete wavelet transform uses filter banks for the construction of the multiresolution time-frequency plane. The Discrete Wavelet Transform analyzes the signal at different frequency bands with different resolutions by decomposing the signal into an approximation and detail information. The decomposition of the signal into different frequency bands obtained by successive high pass g[n] and low pass h[n] filtering of the time domain signal. The combination of high pass g[n] and low pass filter h[n] comprise a pair of analyzing filters. The output of each filter contains half the frequency content, but an equal amount of samples as the input signal. The two outputs together contain the same frequency content as the input signal; however the amount of data is doubled. Therefore down sampling by a factor two, denoted by ↓ 2, is applied to the outputs of the filters in the analysis bank.

Reconstruction of the original signal is possible using the synthesis filter bank. In the synthesis bank the signals are up sampled (↑ 2) and passed through the filters g[n] and h[n]. The filters in the synthesis bank are based on the filters in the analysis bank. Proper choice of the combination of the analyzing filters and synthesizing filters will provide perfect reconstruction. Perfect reconstruction is defined by the output which is generally an estimate of the input, being exactly equal to the input applied. The decomposition process can be iterated with successive approximations being decomposed in return, so that one signal is broken down into many lower-resolution components. Decomposition can be performed as ones requirement.

The Two-Dimensional DWT (2D-DWT) is a multi level decomposition technique. It converts images from spatial domain to frequency domain. One-level of wavelet decomposition produces four filtered and sub-sampled images, referred to as sub bands. The subband image decomposition using wavelet transform has a lot of advantages. Generally, it profits analysis for non-stationary image signal and has high compression rate. And its transform field is represented multiresolution like human's visual system so that can progressively transmit data in low transmission rate line. DWT processes data on a variable time-frequency plane that matches progressively the lower frequency components to coarser time resolutions and the high-frequency components to finer time resolutions, thus achieving a multiresolution analysis. The Discrete Wavelet Transform has become powerful tool in a wide range of applications including image/video processing, numerical analysis and telecommunication. The advantage of DWT over existing transforms, such as discrete Fourier transform (DFT) and DCT, is that the DWT performs a multiresolution analysis of a signal with localization in both time and frequency domain.

## 3. PROPOSED DESIGN

The block diagram of the proposed design is shown in figure 1. It consists of a DWT processor and a pair of external dual-port memories. The two memories are initialized with the pixel values of a gray scale image. In this proposed design, input is provided to the DWT processor by importing an image from the workspace in Matlab. The DWT processor includes DWT filter, memory controller and crossbars. The crossbars are used for interleaving the image pixels i.e. the output of the high pass and low pass filter will be distributed alternatively to the two memory banks.

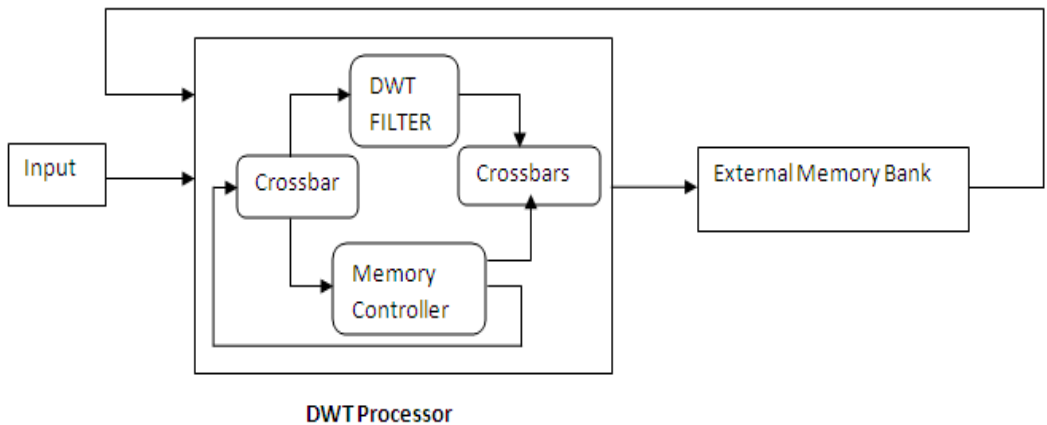

Figure1. Block diagram of proposed design

The DWT filter is designed using discrete wavelet transform. The Discrete Wavelet Transform can be implemented using high pass and low pass filters. The high pass and low pass filters are designed using following transformations:

$$H(2n+1) = X(2n+1) - \operatorname{floor}([X(2n) + X(2n+2)]/2) \ldots (1)$$

$$L(2n) = X(2n) + floor\ ([H(2n-1) + H(2n+1) + 2]/4)....(2)$$

Transformations are performed on each pixel using these filters and this is done as per line basis where lines are defined by start-of-line (sol) and end-of-line (eol). The high pass and low pass filters decompose the image into detail and approximate information respectively. The detail information is basically low scale, high frequency components of the image and it imparts nuance. Whereas the approximate information is high scale, low frequency components of the image and it impart the important part of the image. In the high pass and low pass filter, the new inputs are accepted at one end before previously accepted inputs appear as outputs at the other end. This process is known as pipelining which helps to enhance the speed of the processor. The output of the H and L filters will be alternately distributed to the two memory banks. The data on the 'H' outputs are delayed by 32 cycles relative to the 'L' outputs. Without this delay, the data being written from the 'H' and the 'L' filters would always be trying to write to the same memory bank. With the delay added, they end up always writing to opposite banks.

A memory controller performs the read and writes operation simultaneously. It does not account for latency of getting data from memory or latency of the filter. The memory control signals are all derived from two free-running counters. The reset holds the counts at zero until a start pulse arrives. The bulk of control is determined on per phase basis from the master counter. The state register defines the number of phases. The address logic is derived by recombination of bits from the master counter for each phase. In fact, the read addresses are just the count value -- i.e. the memory read for this phase is just a stride 1 loop through the whole memory bank. The write addresses for this phase repeat each address twice.

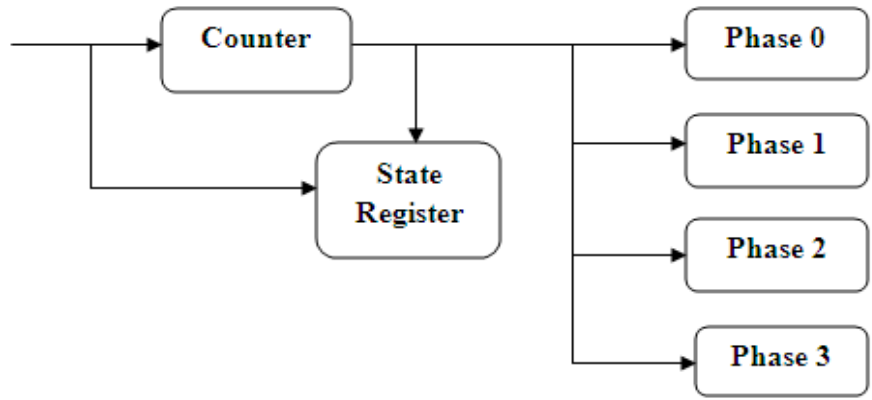

Figure2. Memory Controller

The external memory bank where the write enable is asserted into variable selector block. The variable selector extracts a subset of rows from the input and fed the output to $P_1$ and $P_2$. These products $P_1$ and $P_2$ perform division and multiplication of its inputs and pass it through write inserter. The write inserter passes first or third input based on the value of second input and output is fed to the read section. This means one word is inserted to the specific address location of external memory bank. And the read section picks up the appropriate word from the memory vector. In case of overlapping of address, the read is done before the write changes the stored word.

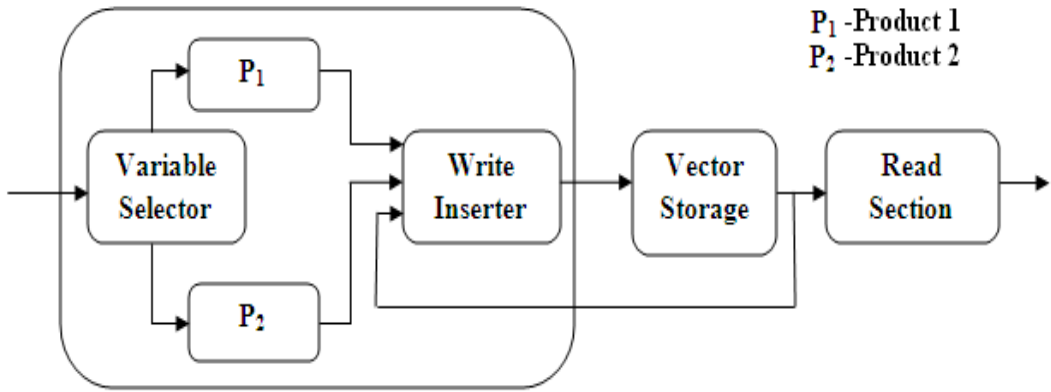

Figure3. External Memory Bank

## 4. HARDWARE IMPLEMENTATION RESULTS

The proposed model has designed and simulated using Simulink and Xilinx System Generator block sets. The simulated has been accomplished by using DWT filter in the proposed model. The DWT filter uses high pass and low pass filter to decompose the image into its detail and approximate information respectively. The decomposition of the image is shown in the figure 4.

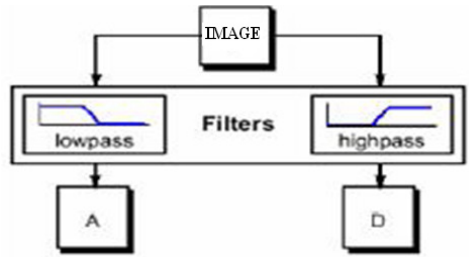

Figure 4. Decomposition of the Image

The decomposition process can be iterated with successive approximations being decomposed into many lower resolution components. This is also called as the wavelet decomposition tree. The iteration of the decomposition process is shown in figure 5.

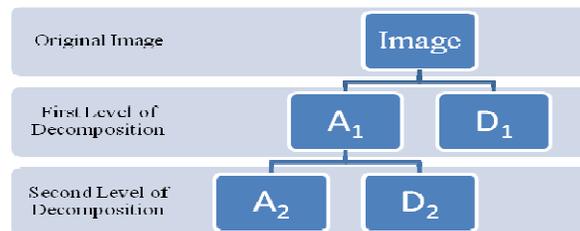

Figure 5 Iteration of Decomposition Process

2D-DWT is applied on grayscale image which is shown in figure 6. It transforms an image into sub-bands such that the wavelet coefficients in the lower level sub-bands typically contain more energy than those in higher level sub-bands.

26

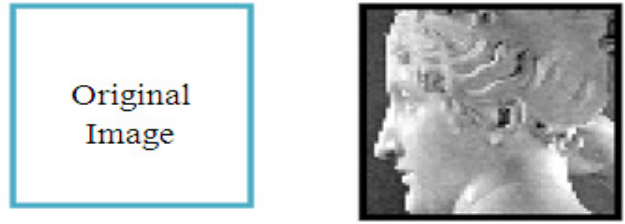

Figure 6. Original Image

It can be accomplished by applying one-dimensional DWT filter in a separable manner. The first stage of the DWT divides an image into four sub-bands by applying low-pass and high pass filters. The first level of decomposition is consists of two steps. In the first step, each row of an image is transformed using a 1D vertical analysis filter bank. The first step is shown in figure 7.

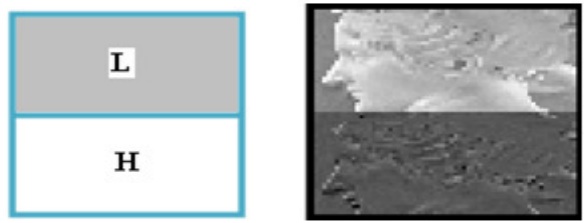

Figure 7. Each row replaced by 1DWT

In the second step of the first level of decomposition, each column of the transformed image is again transformed using same filter bank horizontally. The second step is shown in figure 8. Thus first level of decomposition produces four filtered and sub-sampled images.

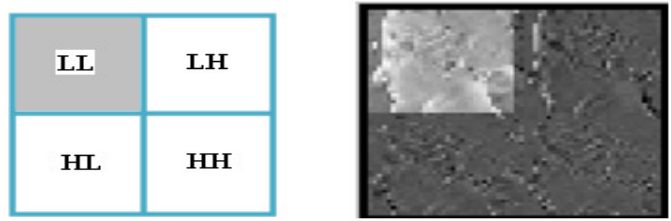

Figure 8. Each row replaced by 1DWT

For the second level of decomposition, DWT further divides the lowest sub-band using the same filtering method as above. The lowest sub-band has been decomposed into further four sub-bands. Each row and column of the lowest sub-band has been replaced by 1D-DWT. The result of the second level of decomposition has been shown in figure 9.

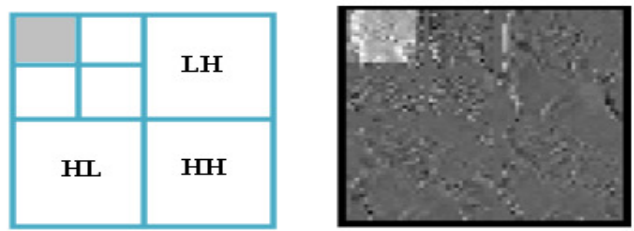

Figure 9. Second level of Decomposition

The developed VHDL code has been simulated using Modelsim, synthesized using Xilinx Synthesizer tool (XST) and implemented on Spartan 2 based XC2S100-5tq144 FPGA and Spartan 3E based XC3S500E-4fg320 FPGA device. The resource utilization of Spartan 2 and 3E based FPGA device has been shown in table 1. It can be observed from the table 1 that Spartan 3E based FPGA target device uses less number of resources than Spartan 2 based FPGA device which helps to enhance speed along with optimal utilization.

Table 1 Device Utilization Based on Spartan 2 and 3 FPGA

| Device Utilization Summary | | | | | | |
|---|---|---|---|---|---|---|
| Logic Utilization | Used | | Available | | Utilization | |
| | Spartan 2 | Spartan 3 | Spartan 2 | Spartan 3 | Spartan 2 | Spartan 3 |
| Number of Slice | 278 | 272 | 1200 | 4656 | 23% | 5% |
| Number of Slice Flip Flops | 491 | 498 | 2400 | 9312 | 20% | 5% |
| Number of 4input LUTs | 363 | 334 | 2400 | 9312 | 15% | 3% |
| Number of bonded IOBs | 91 | 91 | 92 | 232 | 98% | 39% |
| Number of GCLKs | 1 | 1 | 4 | 24 | 25% | 4% |

The timing summary of Spartan 2 and 3 based on XC2S100-5tq144 and XC3S500E-4fg320 FPGA device, respectively, has been shown in table 2. The proposed design implemented on Spartan 2 based FPGA can work at maximum operating frequency of 143 MHz. And when the proposed design implemented on Spartan 3E based FPGA can work at maximum operating frequency of 231.26 MHz by utilizing considerable fewer resources on the target device in terms of LUTS, Flip flops and slices.

Table 2 Timing Summary Based on Spartan 2 and 3 FPGA

| Description | Speed | | | |
|---|---|---|---|---|
| | Spartan 2 | | Spartan 3 | |
| | (ns) | (MHz) | (ns) | (MHz) |
| Minimum Period | 6.991 | 143.041 | 4.324 | 231.26 |
| Minimum input arrival time before clock | 2.827 | 353.73 | 1.946 | 513.87 |
| Maximum ouput required time after clock | 8.189 | 122.11 | 4.310 | 232.01 |

The total power consumption of the proposed design based on XC3S500E-4fg320 FPGA device has been calculated using XPower utility. It can be observed from the table 5 that proposed design has consumed 117mW at 28 degree C.

Table 3 Power Consumption

| Name | Value | Used | Total Available | Utilization (%) |
|---|---|---|---|---|
| Clocks | 0.01036 (W) | 1 | --- | --- |
| Logic | 0.00069 (W) | 333 | 9312 | 3.6 |
| Signals | 0.00187 (W) | 569 | --- | --- |
| IOs | 0.02350 (W) | 91 | 232 | 39.2 |
| Total Quiescent Power | 0.08144 (W) | | | |
| Total Dynamic Power | 0.03641 (W) | | | |
| Total Power | 0.11786 (W) | | | |
| Junction Temp | 28.1 (degrees C) | | | |

Finally proposed design results based on Spartan 2 & 3E FPGA have been compared with [8]. The proposed design has shown an improvement, in table 4, of 15% in speed as compared to [8] and spartan 2 based design by consuming considerable less number of slices available on target FPGA device.

Table 4 Performance Comparison with existing models

| Device Utilization Summary | DWT/IDWT Prototype Design | Proposed Design | |
|---|---|---|---|
| | VirtexII | SpartanII | Spartan3E |
| Number of Slices | 1907 out of 9280 | 278 out of 1200 | 272 out of 4656 |
| Minimum period | 4.973 ns | 6.991 ns | 4.324 ns |
| Maximum frequency | 201.092 MHz | 143.041 MHz | 231.267 MHz |
| Total estimated power consumption | 861 mW | 429.93 mW | 117.86 mW |

## 5. CONCLUSIONS

In this paper, high speed and area efficient DWT processor based Image Compression model has been presented. The pipelined partially serial architecture is introduced to enhance the speed and area efficiency. The proposed design can operate at a maximum frequency of 231 MHz by consuming of 117mW power at 28°C junction temperature. An improvement of 15% in speed has been achieved by consuming considerably less number of resources of Spartan 3E based XC3S500E-4fg320 FPGA device to provide cost effective solutions for real time image processing applications.

**Authors**

**Sugreev Kaur:** Miss Sugreev Kaur is *currently pursuing M.E. degree from National Institute of Technical Teachers' Training and Research, Chandigarh. She has completed B.Tech degree in Electronics and Communication from PTU, Jalandhar, Punjab, in 2007. Miss Sugreev Kaur has authored a paper in national and international conference. Miss Sugreev's interest areas are VLSI design and Embedded System design.*

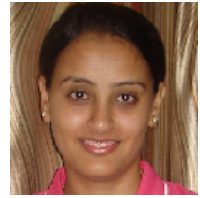

**Rajesh Mehra:** *Mr. Rajesh Mehra is currently Assistant Professor at National Institute of Technical Teachers' Training & Research, Chandigarh, India. He is pursuing his PhD from Punjab University, Chandigarh, India. He has completed his M.E. from NITTTR, Chandigarh, India and B.Tech. from NIT, Jalandhar, India. Mr. Mehra has 14 years of academic experience. He has authored more than 30 research papers in national, international conferences and reputed journals. Mr. Mehra's interest areas are VLSI Design, Embedded System Design, Advanced Digital Signal Processing, Wireless & Mobile Communication and Digital System Design. Mr. Mehra is life member of ISTE*

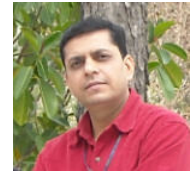